\documentclass[11pt]{article}
\usepackage{graphicx}
\usepackage{amsmath,amssymb}
\def\1{{\bf 1}}

\def\be{\begin{equation}}
\def\ee{\end{equation}}

\begin{document}
\noindent {\it Please cite this paper as {\sl ``Zhang JP, The Lennard-Jones Potential Minimization Problem for Prion AGAAAAGA Amyloid Fibril Molecular Modeling, arXiv:submit/0261672 [math-ph] 7 Jun 2011, in prepare to submit to Journal."} and welcome to together study 3nhc.pdb, 3nve.pdb, 3nvf.pdb, 3nvg.pdb and 3nvh.pdb of PDB bank with Dr JP Zhang.}\\

\noindent {\large \bf The Lennard-Jones Potential Minimization Problem for Prion AGAAAAGA Amyloid Fibril Molecular Modeling}\\

\bigskip

\noindent {\Large Jiapu Zhang}\\

\noindent Centre in Informatics and Applied Optimization \& Graduate School of ITMS,  University of Ballarat, Mount Helen, VIC 3353, Australia.
Emails: j.zhang@ballarat.edu.au, jiapu\_zhang@hotmail.com, Phones: 61-3 5327 9809, 61-423 487 360 (mobile).\\

\noindent {\bf Abstract:} 
The simplified Lennard-Jones (LJ) potential minimization problem is
\begin{equation*}
\mbox{minimize}~~~f(x)=4\sum_{i=1}^N \sum_{j=1,j<i}^N \left(
\frac{1}{\tau_{ij}^6} -\frac{1}{\tau_{ij}^3}
\right)~~~\mbox{subject to}~~~ x\in \mathbb{R}^n,
\end{equation*}
\noindent where $\tau_{ij}=(x_{3i-2}-x_{3j-2})^2
                +(x_{3i-1}-x_{3j-1})^2
                +(x_{3i}  -x_{3j}  )^2$,
$(x_{3i-2},x_{3i-1},x_{3i})$ is the
coordinates of atom $i$ in $\mathbb{R}^3$, $i,j=1,2,\dots,N(\geq 2 \quad \text{integer})$, $n=3N$ and $N$ is the whole number of atoms. The nonconvexity of the objective function and the huge number of local minima, which is growing exponentially with $N$, interest many mathematical optimization experts. The global minimizer should be just at the point of the bottom of the LJ potential well. Based on this point, this paper tackles this problem illuminated by amyloid fibril molecular model building. The 3nhc.pdb, 3nve.pdb, 3nvf.pdb, 3nvg.pdb and 3nvh.pdb of PDB bank are used for the successful molecular modeling.\\


\section{At the Bottom of the L-J Potential Well}
\begin{figure}[h!]
\centerline{
\includegraphics[width=4.2in]{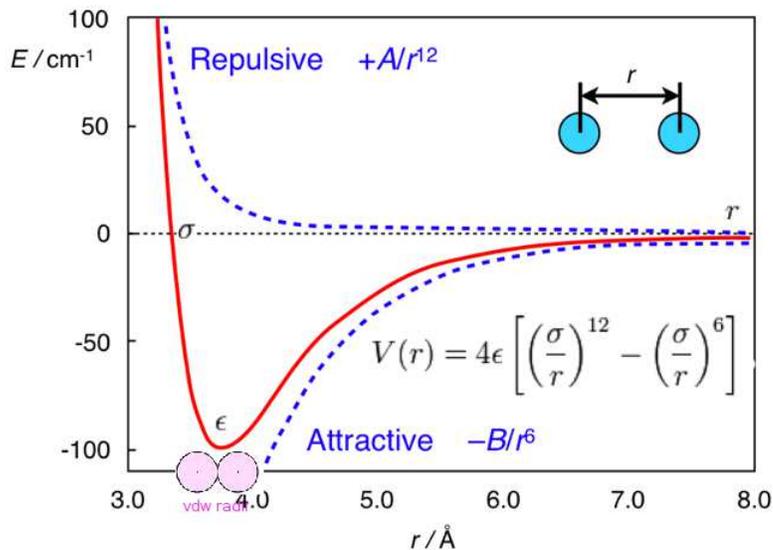}
}
\caption{The Lennard-Jone Potential (formulas (\ref{LJ_r_form}) and (\ref{LJ_AB_form})) (This Figure can be found in website {\tiny http://homepage.mac.com/swain/CMC/DDResources/mol\_interactions/molecular\_interactions.html}}). \label{LJ_potential}
\end{figure}

Neutral atoms are subject to two distinct forces in the limit of large distance and short distance: a dispersion force (i.e. attractive van der Waals (vdw) force) at long ranges, and a repulsion force, the result of overlapping electron orbitals. The Lennard-Jones (L-J) potential represents this behavior ({\small http://en.wikipedia.org/wiki/Lennard-Jones\_potential}, or \cite{locatelli2008} and references therein). The L-J potential is of the form
\begin{equation} \label{LJ_r_form}
V(r)=4\varepsilon \left[ (\frac{\sigma}{r})^{12} - (\frac{\sigma}{r})^6 \right],
\end{equation}
where $r$ is the distance between two atoms, $\varepsilon$ is the depth of the potential well and $\sigma$ is the atom diameter; these parameters can be fitted to reproduce experimental data or deduced from results of accurate quantum chemistry calculations. The $(\frac{\sigma}{r})^{12}$ term describes repulsion and the $(\frac{\sigma}{r})^6$ term describes attraction (Fig. \ref{LJ_potential}). 
In Fig. \ref{LJ_potential} we may see two points: (I) $V(r)=0$ (but the value of $V(r)$ is not the minimal value) when $r=\sigma$ (i.e. the distance between two atoms equals to the sum of {\it atom radii} of the atoms); and (II) when $r=2^{1/6}\sigma$ (i.e. the distance between two atoms equals to the sum of {\it vdw radii} of the atoms), the value of $V(r)$ reaches its minimal value $-\varepsilon$ (i.e. the bottom of the potential well; the force between the atoms is zero at this point). This paper is written based on (II). If we introduce the coordinates of the atoms whose number is denoted by $N$ and let $\varepsilon = \sigma =1$ be the reduced units, the form (\ref{LJ_r_form}) becomes
\begin{equation}\label{LJ_x_form}
f(x)=4\sum_{i=1}^N \sum_{j=1,j<i}^N \left( \frac{1}{\tau_{ij}^6}
-\frac{1}{\tau_{ij}^3} \right),
\end{equation}
\noindent where $\tau_{ij}=(x_{3i-2}-x_{3j-2})^2
                          +(x_{3i-1}-x_{3j-1})^2
                          +(x_{3i}  -x_{3j}  )^2 =||X_i-X_j||^2_2$
and $(x_{3i-2},x_{3i-1},x_{3i})$ is the coordinates of atom $i$, $i,j=1,2,\dots, N (\geq 2)$. The minimization of L-J potential $f(x)$ on $\mathbb{R}^n$ (where $n=3N$) is an optimization problem:
\begin{equation}\label{LJ_f_form}
\min_{s.t. x\in \mathbb{R}^{3N}} f(x).
\end{equation}\\

For (\ref{LJ_f_form}), when its global optimization solution is reached, the value $r$ in (\ref{LJ_r_form}) should be the sum of two {\it vdw radii} of the two atoms interacted. The three dimensional structure of a molecule with $N$ atoms can be described by specifying the 3-Dimensional coordinate positions $X_1, X_2, \dots, X_N \in \mathbb{R}^3$ of all its atoms. Given bond lengths $r_{ij}$ between a subset $S$ of the atom pairs, the determination of the molecular structure is
\begin{eqnarray}
(\mathcal{P}_0 ) \quad to \quad find \quad &X_1,X_2,\dots ,X_N  \quad s.t. \quad ||X_i-X_j||=r_{ij}, (i,j)\in S,  \label{orginal_problem}
\end{eqnarray}
where $||\cdot ||$ denotes a norm in a real vector space and it is calculated as the Euclidean distance 2-norm in this paper. (\ref{orginal_problem}) can be reformulated as a mathematical global optimization problem (GOP)
\begin{eqnarray}
(\mathcal{P} ) \quad &\min P(X)=\sum_{(i,j)\in S} w_{ij} (||X_i-X_j||^2 -r_{ij}^2 )^2  \label{prime_problem}
\end{eqnarray}
in the terms of finding the global minimum of the function $P(X)$, where $w_{ij}, (i,j)\in S$ are positive weights, $X = (X_1, X_2, \dots, X_N)^T \in \mathbb{R}^{N\times 3}$ \cite{morew1997} and usually $S$ has many fewer than $N^2/2$ elements due to the error in the theoretical or experimental data \cite{zoubs1997,grossols2009}. There may even not exist any solution $X_1, X_2, \dots, X_N$ to satisfy the distance constraints in (\ref{orginal_problem}), for example when data for atoms $i, j, k \in S$ violate the triangle inequality; in this case, we may add a perturbation term $-\epsilon^TX$ to $P(X)$:
\begin{eqnarray}
(\mathcal{P}_{\epsilon} ) \quad &\min P_{\epsilon}(X)=\sum_{(i,j)\in S} w_{ij} (||X_i-X_j||^2 -r_{ij}^2 )^2 -\epsilon^TX, \label{prime_approx_problem}
\end{eqnarray}
where $\epsilon \geq 0$. Thus, the L-J potential optimization problem (\ref{LJ_f_form}) is rewritten into the optimization problem (\ref{prime_approx_problem}). In this paper we elegantly solve \ref{prime_approx_problem} (thus solve (\ref{LJ_f_form})) illuminated by the prion AGAAAAGA amyloid fibril molecular modeling.\\

\section{L-J Potential in Prion AGAAAAGA Amyloid Fibril Molecular Modelling}
In 2007, Sawaya et al. got a breakthrough finding: the atomic structures of all amyloid fibrils revealed steric zippers, with strong vdw interactions (LJ) between $\beta$-sheets and hydrogen bonds (HBs) to maintain the $\beta$-strands \cite{sawaya2007}. Similarly as (\ref{LJ_r_form}), i.e. the potential energy for the vdw interactions (Fig. \ref{LJ_potential}) between $\beta$-sheets:
\begin{equation} \label{LJ_AB_form}
V_{LJ}(r)=\frac{A}{r^{12}} -\frac{B}{r^6},
\end{equation}
the potential energy for the HBs between the $\beta$-strands has a similar formula 
\begin{equation} \label{HB_r_form}
V_{HB}(r)= \frac{C}{r^{12}} -\frac{D}{r^{10}} ,
\end{equation}
where $A,B,C,D$ are constants given. Thus, the amyloid fibril molecular model building problem is reduced to well solve the optimization problem (\ref{LJ_f_form}) or (\ref{prime_approx_problem}).\\

In this section, we will use suitable templates 3nvf.pdb, 3nvg.pdb and 3nvh.pdb from the Protein Data Bank (http://www.rcsb.org/) to build some amyloid fibril models. The models were built by the mutations in the use of SPDBV.4.01.PC (which make all the vdw contacts between the two $\beta$-sheets are very far), any Optimization Solver (which will remove the bad vdw/HB contacts) and Optimization program of Amber 11 (which furthermore refines the models and removes all the bad contacts and relax the models into a perfect way).\\

\begin{figure}[h!]
\centerline{
\includegraphics[scale=0.15]{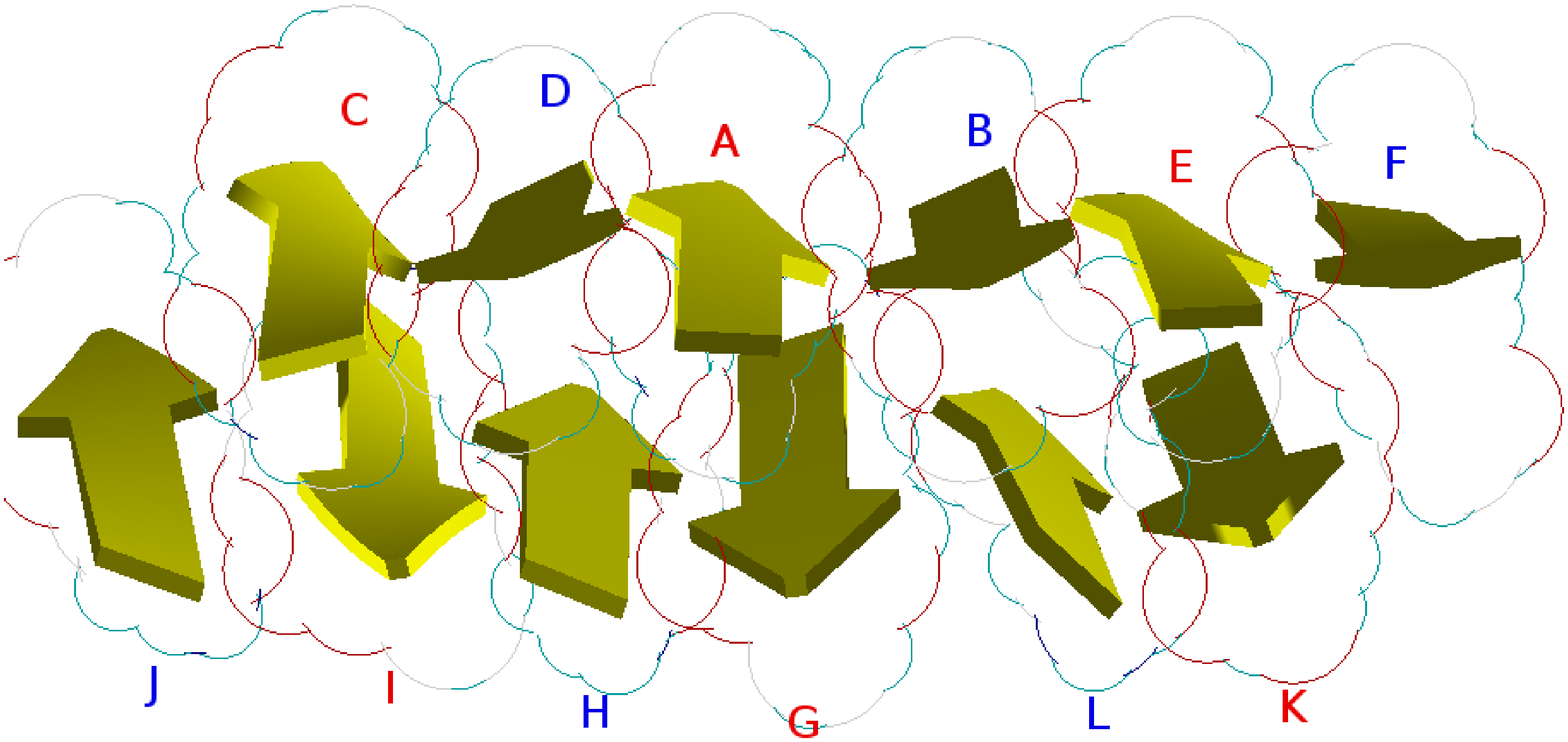}
\includegraphics[scale=0.15]{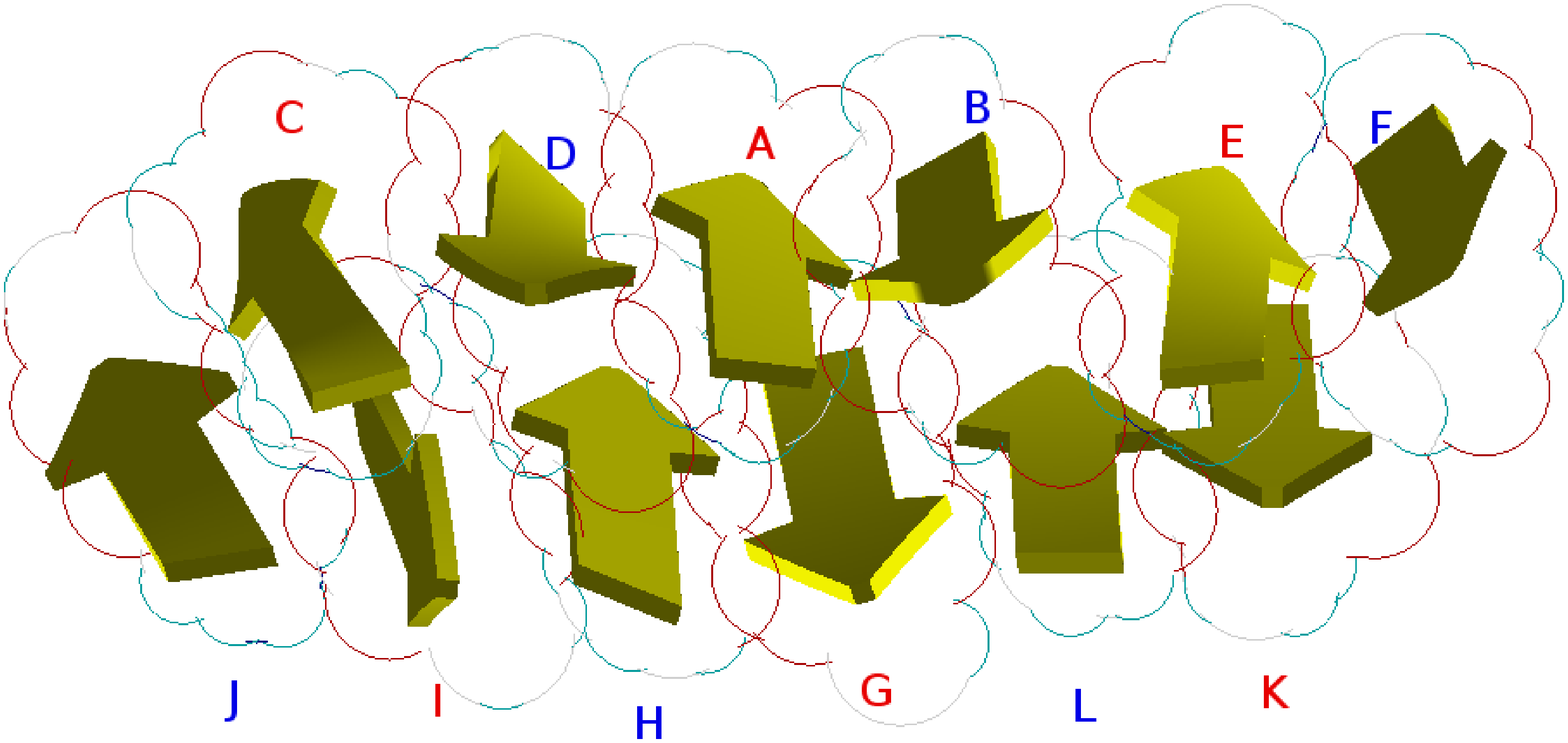}
\includegraphics[scale=0.15]{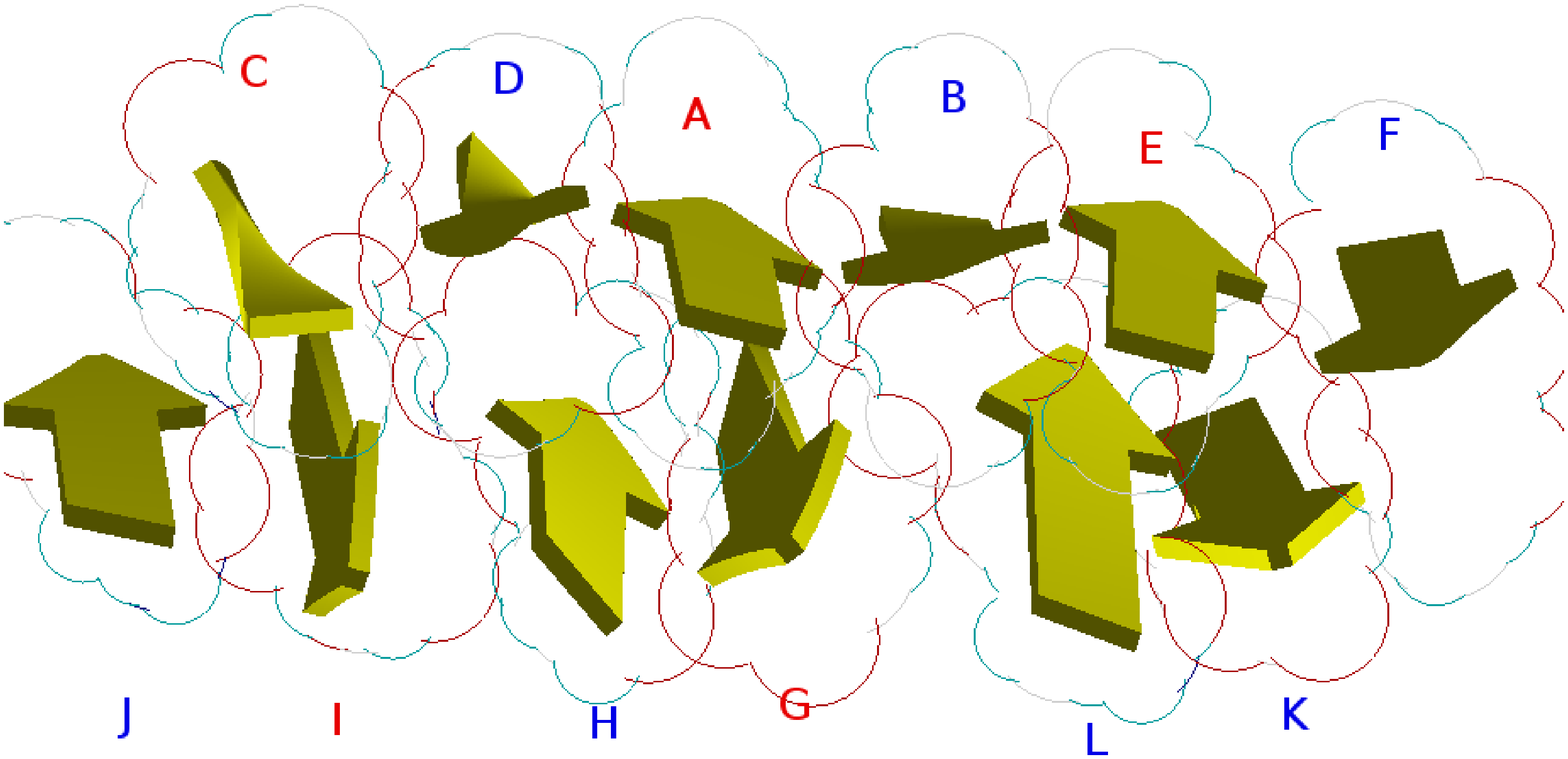}
}
\caption{Perfect 3nhc-Models 1-3 (from left to right respectively) for prion AGAAAAGA segment 113-120.}
\label{3nhc_CDT_models_min2}
\end{figure}

\begin{figure}[h!]
\centerline{
\includegraphics[scale=0.15]{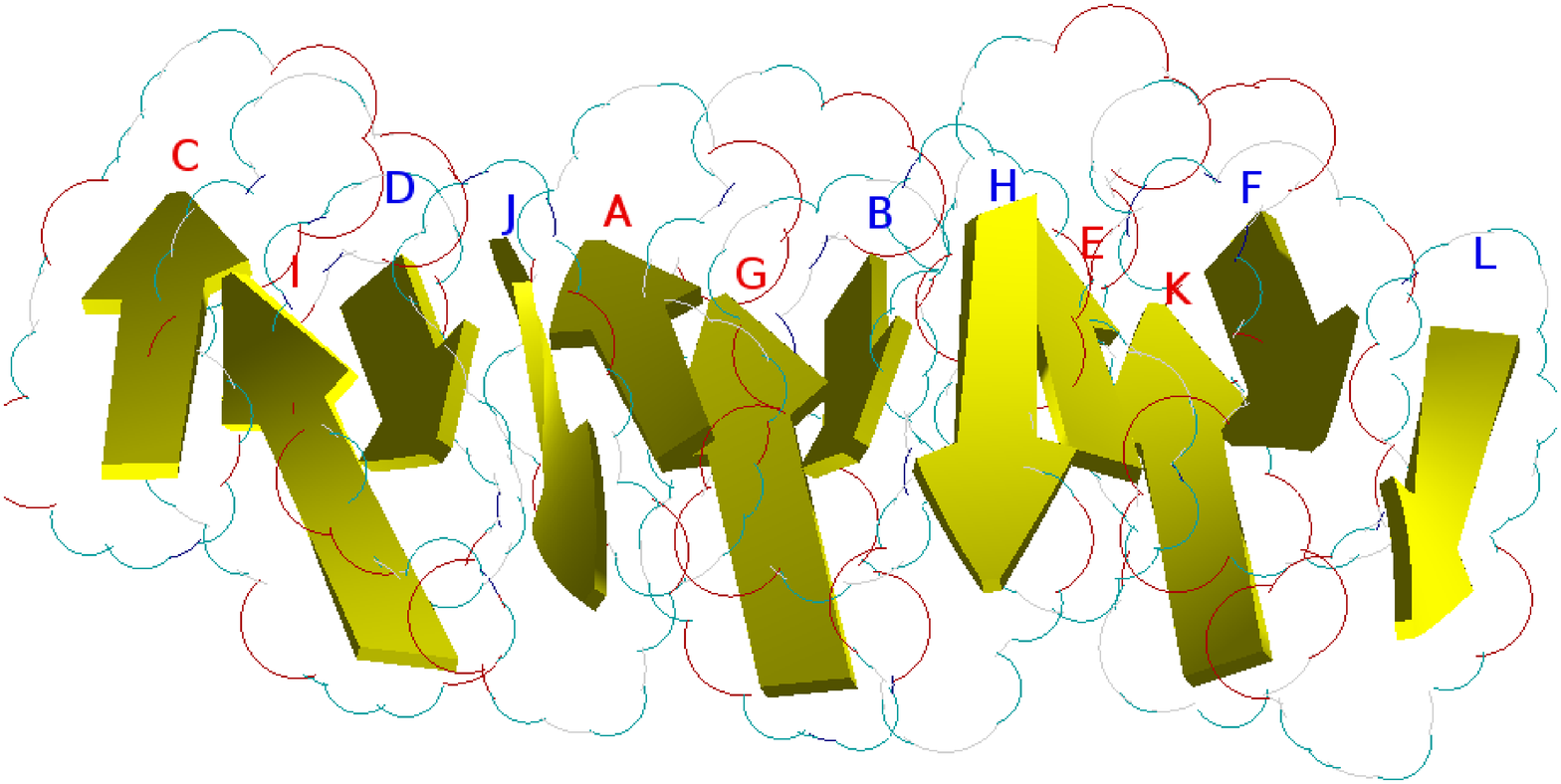}
\includegraphics[scale=0.15]{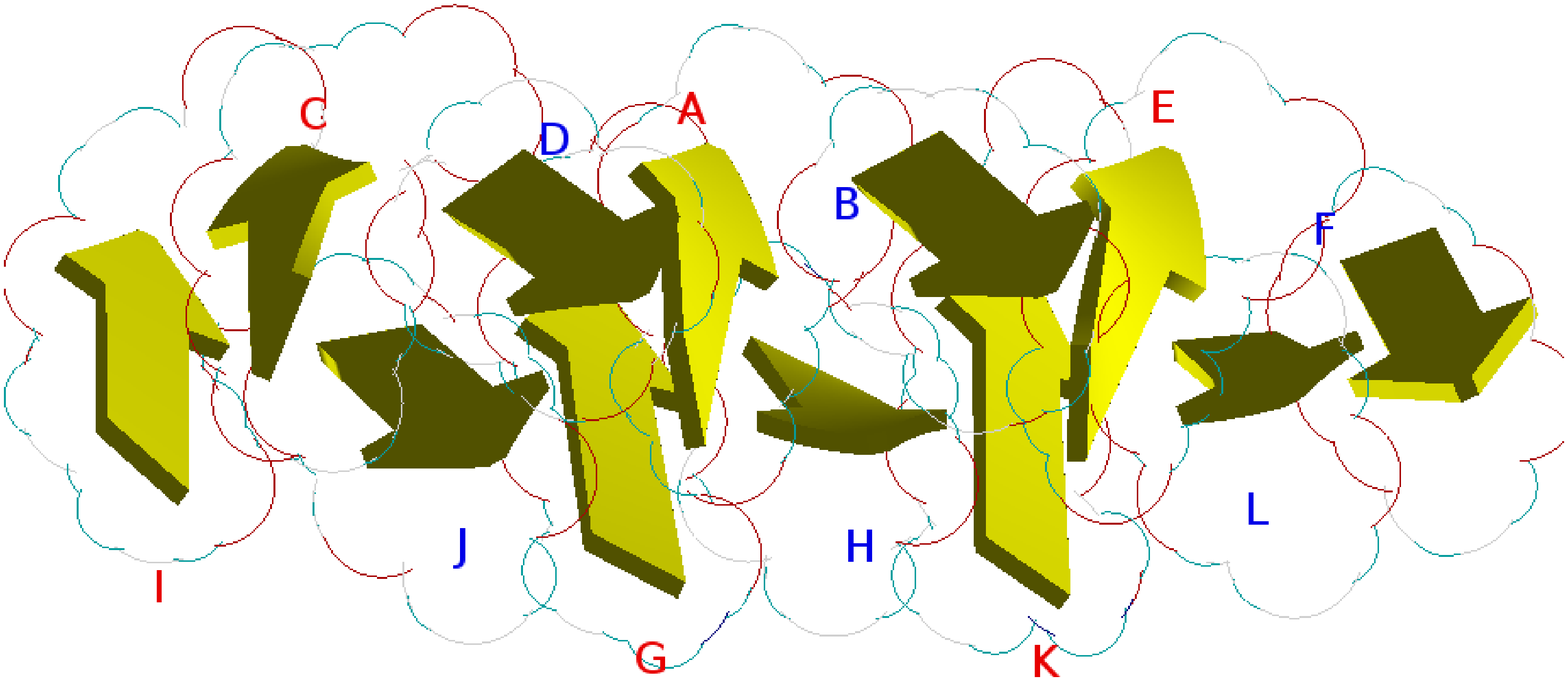}
\includegraphics[scale=0.15]{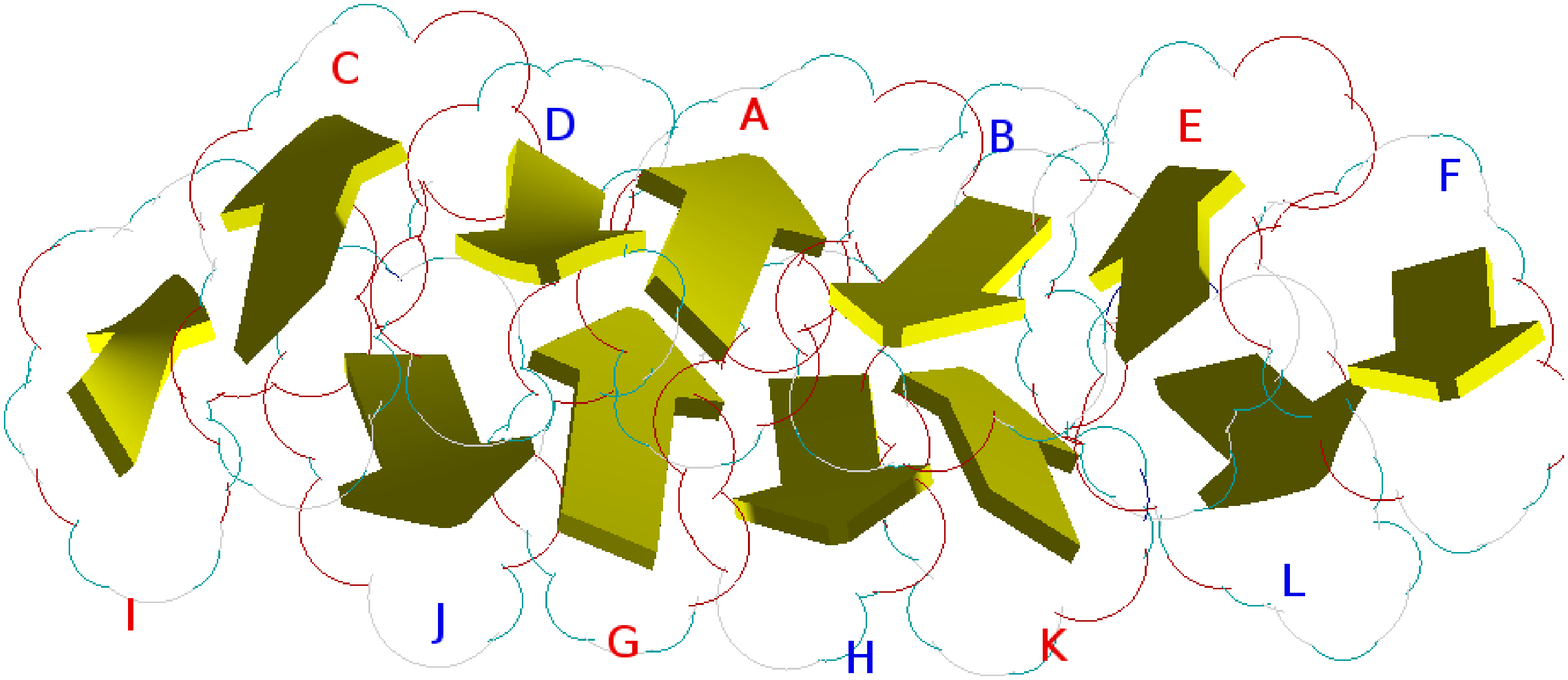}
}
\caption{Perfect 3nve-Models 1-3 (from left to right respectively) for prion AGAAAAGA segment 113-120.}
\label{3nve_CDT_models_min2}
\end{figure}

\begin{figure}[h!]
\centerline{
\includegraphics[scale=0.2]{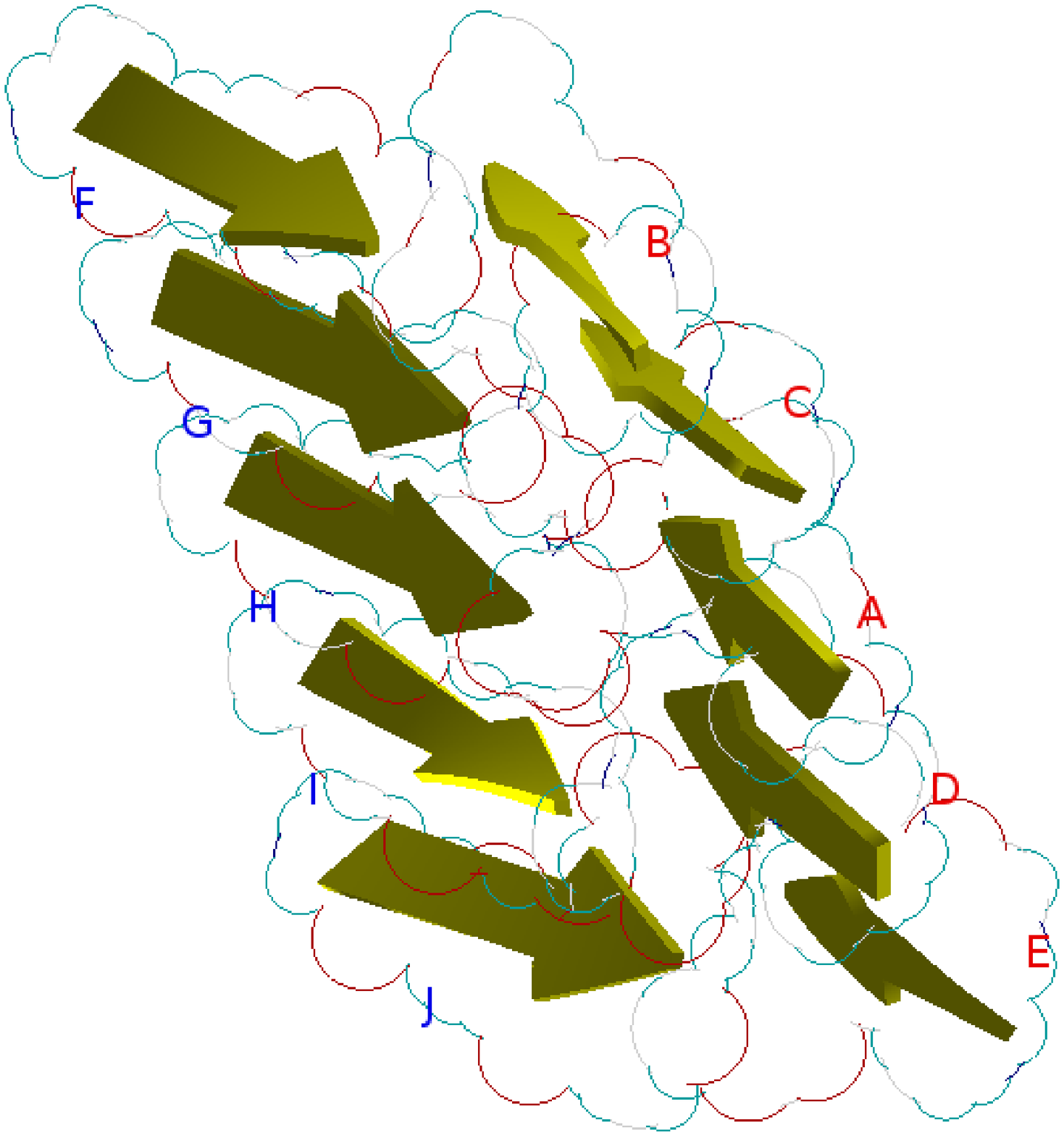}
\includegraphics[scale=0.2]{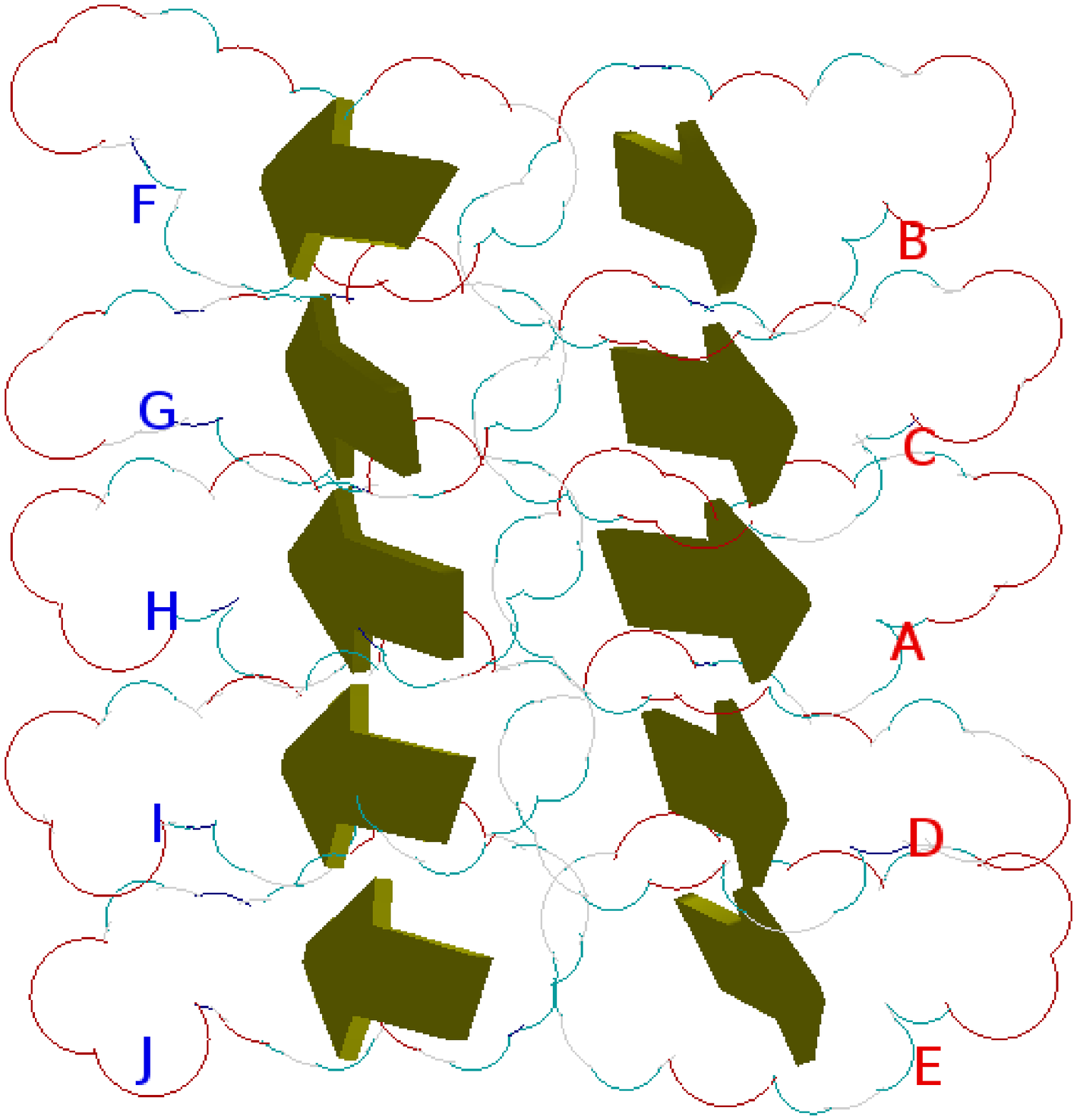}
\includegraphics[scale=0.2]{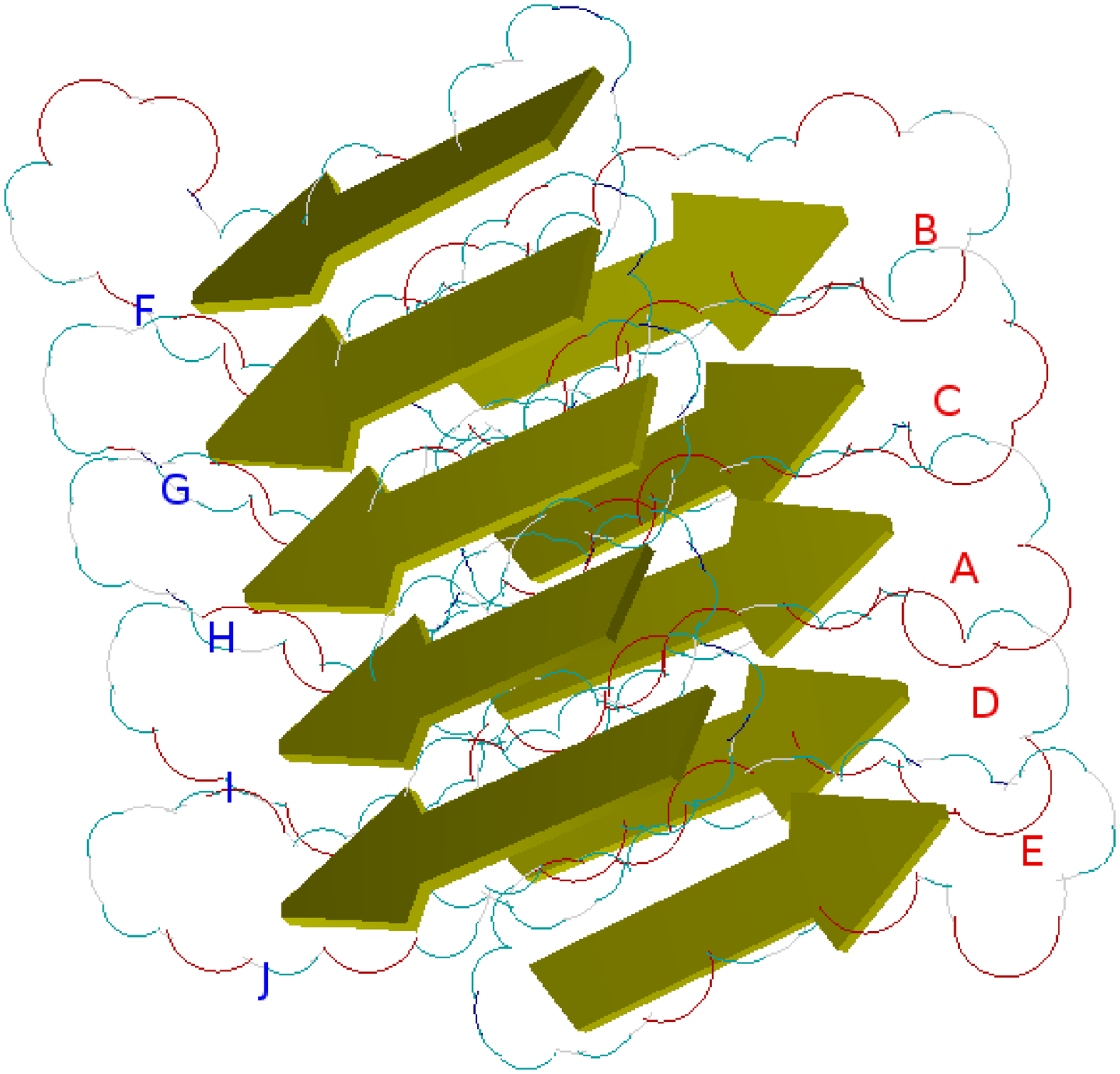}
}
\caption{Perfect 3nvf-Models 1-3 (from left to right respectively) for prion AGAAAAGA segment 113-120.}
\label{3nvf_CDT_models_min2}
\end{figure}

\begin{figure}[h!]
\centerline{
\includegraphics[scale=0.2]{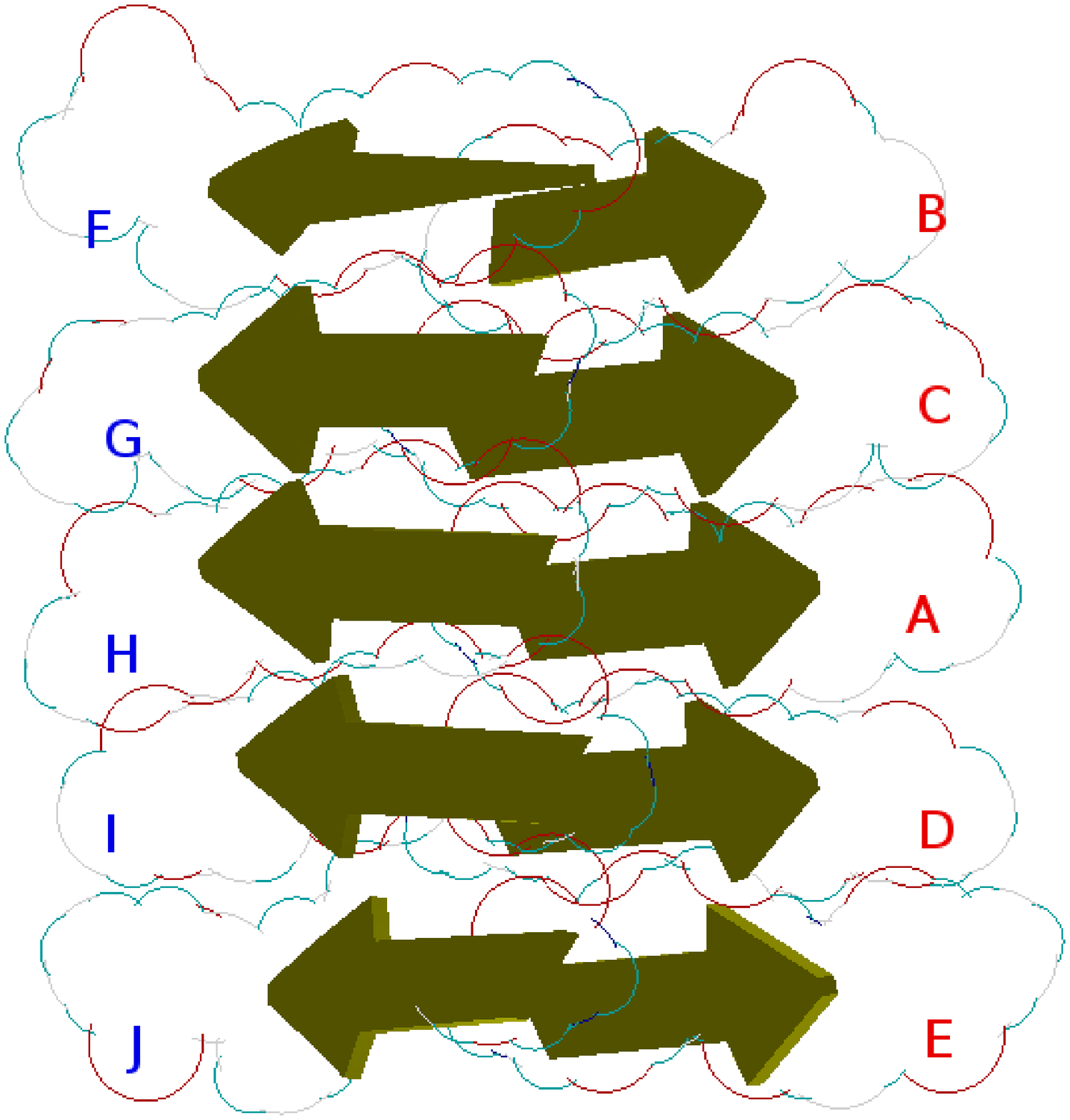}
\includegraphics[scale=0.2]{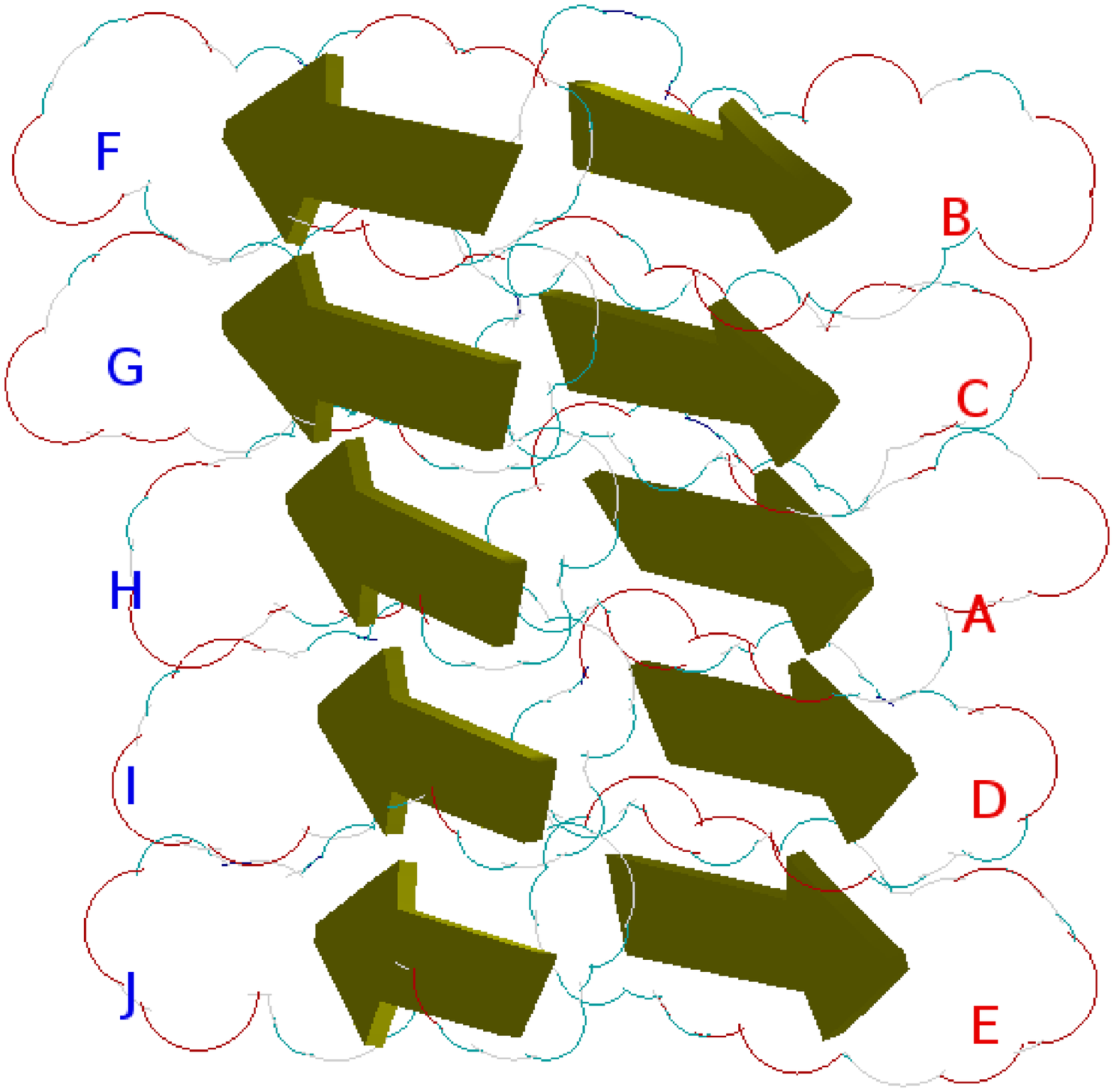}
\includegraphics[scale=0.2]{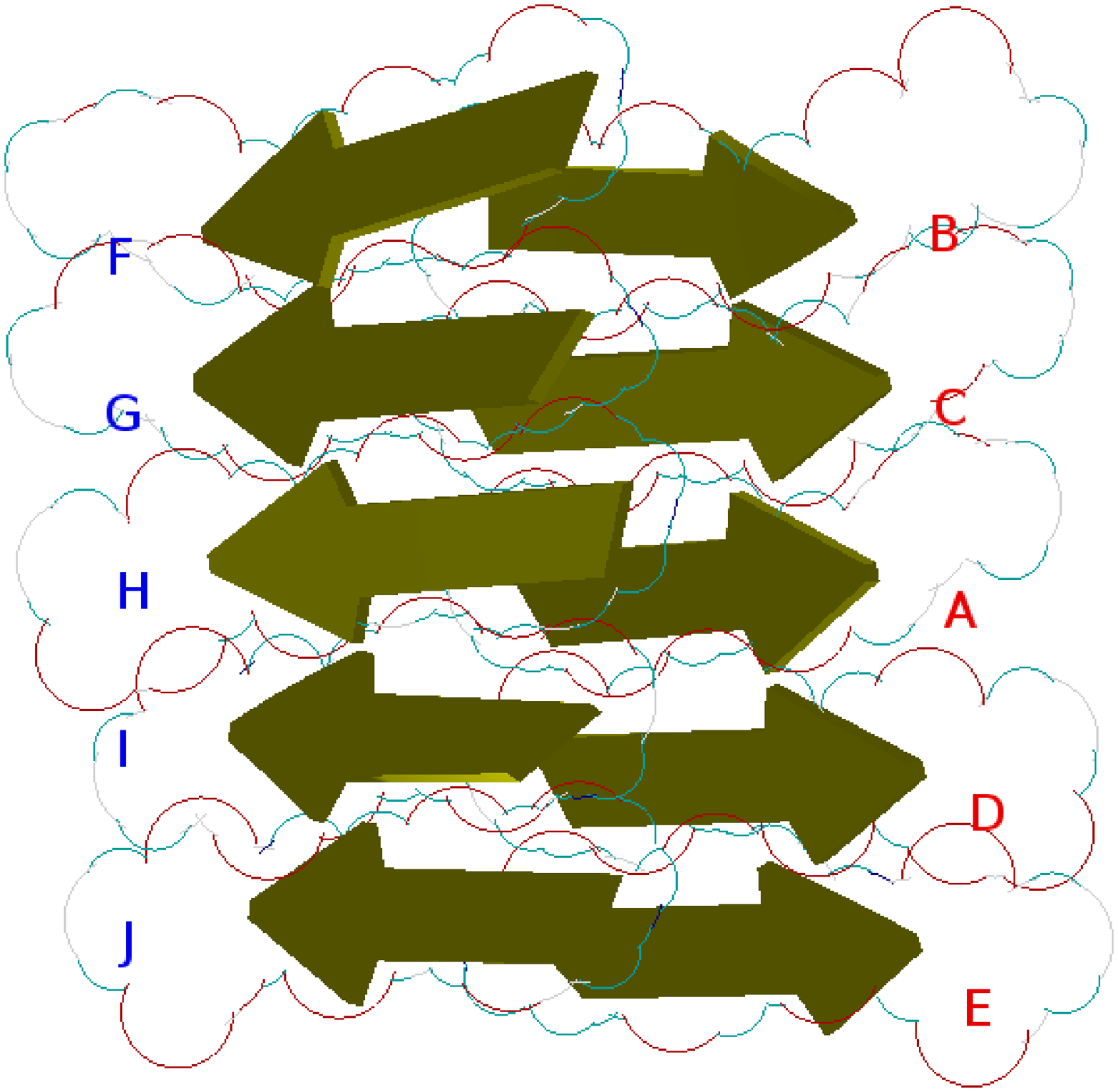}
}
\caption{Perfect 3nvg-Models 1-3 (from left to right respectively) for prion AGAAAAGA segment 113-120.}
\label{3nvg_CDT_models_min2}
\end{figure}

\begin{figure}[h!]
\centerline{
\includegraphics[scale=0.2]{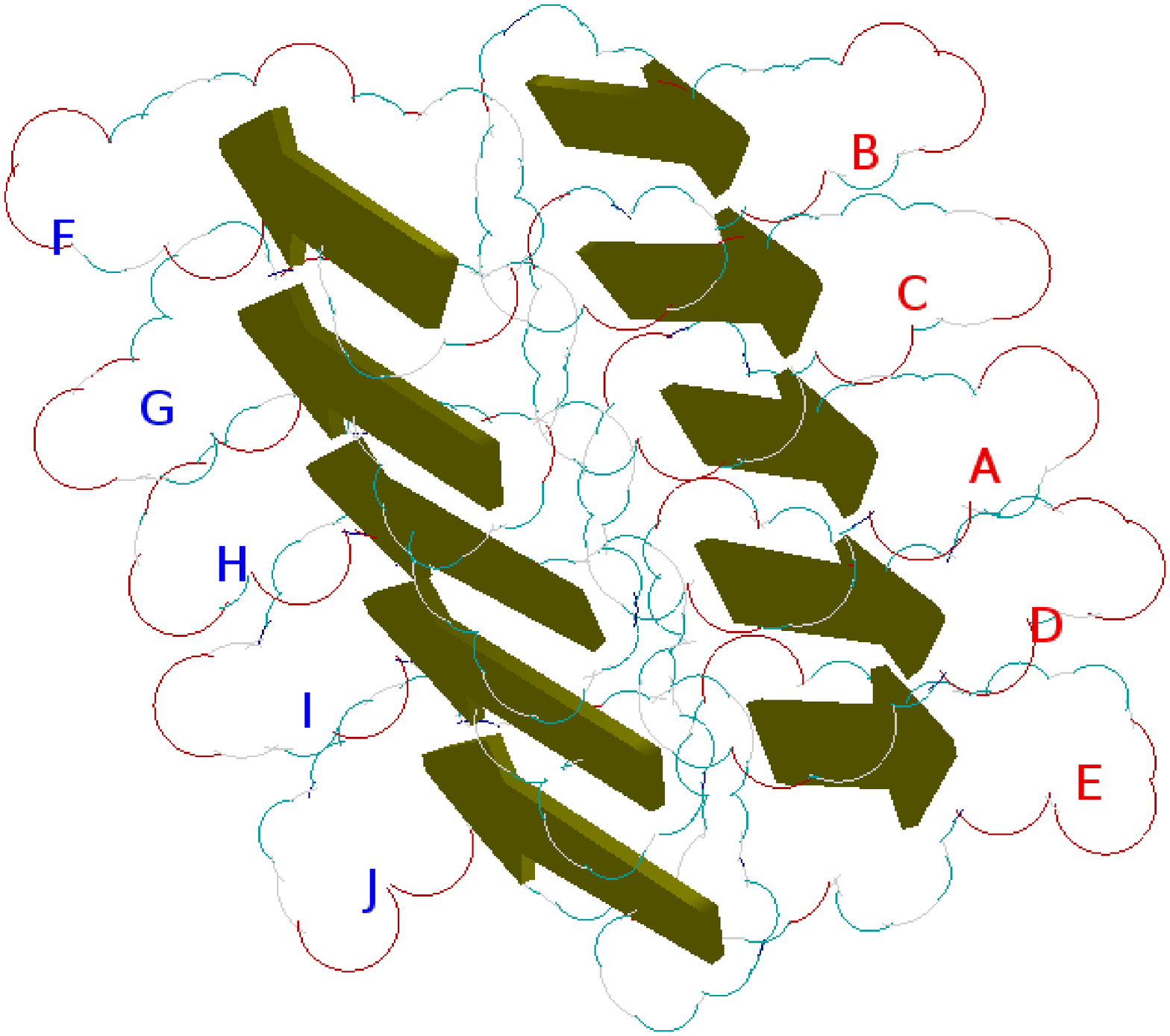}
\includegraphics[scale=0.2]{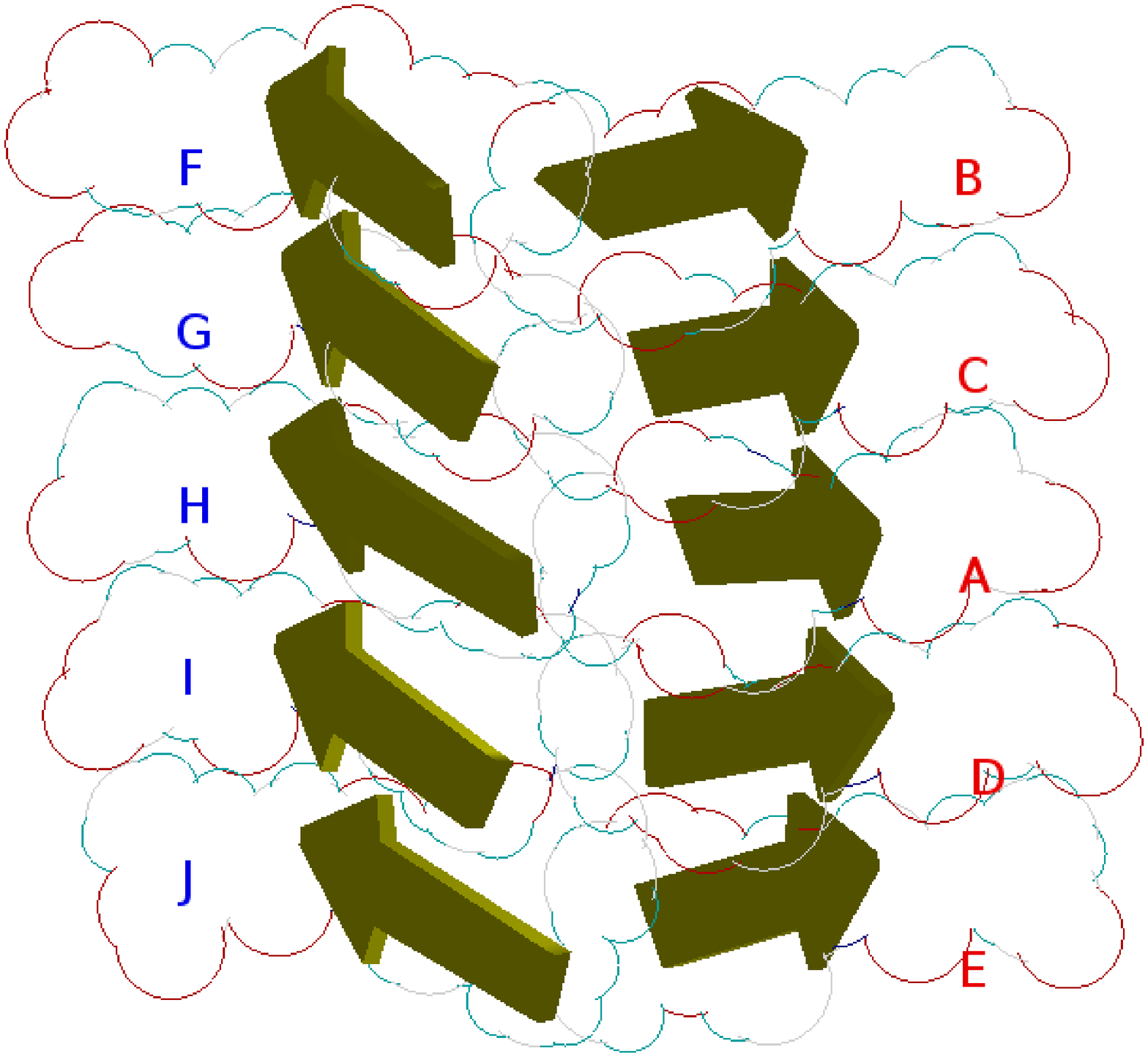}
}
\caption{Perfect 3nvh-Models 1-2 (from left to right respectively) for prion AGAAAAGA segment 113-120.}
\label{3nvh_CDT_models_min2}
\end{figure}

The amyloid fibril models of prion AGAAAAGA segment lastly refined by Amber 11 are illuminated in Fig.s \ref{3nhc_CDT_models_min2}-\ref{3nvh_CDT_models_min2}. All these models are without any bad contact now (checked by package Swiss-PdbViewer), and the vdw interactions between the two $\beta$-sheets are in a very perfect way now.
All the initial structures after SPDBV.01.PC mutations have very far vdw contacts between the two $\beta$-sheets. Now the vdw contacts come closer and reach a state with the lowest potential energy, which has perfect vdw contacts as shown in Fig.s \ref{3nvf_CDT_models_min2}-\ref{3nvh_CDT_models_min2}.\\

\end{document}